\documentclass[a4paper,11pt]{article}

\usepackage[a4paper,margin=2.5cm]{geometry}

\usepackage[utf8]{inputenc}

\usepackage[english]{babel}

\usepackage{amsfonts}

\usepackage{graphicx}

\usepackage{subfigure}

\usepackage[thinlines]{easytable}

\usepackage{natbib}

\setcitestyle{authoryear,round,semicolon}

\usepackage{hyperref}

\usepackage{algorithm}
\usepackage{algpseudocode}

\title{The Surprising Robustness of Partial Least Squares}

\author{João B. Assunção\footnote{ORCID: \url{https://orcid.org/0000-0002-5576-3473}.}\\Pedro Afonso Fernandes\footnote{ORCID: \url{https://orcid.org/0000-0001-5762-5157}. Correspondence: Universidade Católica Portuguesa, Católica Lisbon School of Business \& Economics, Palma de Cima, Building 5, 4th floor, Room 5430, 1649-023 Lisboa, Portugal. Email: paf@ucp.pt}\\\\Universidade Católica Portuguesa\\Católica Lisbon School of Business \& Economics\\Católica Lisbon Research Unit in Business \& Economics (CUBE)\\Católica Lisbon Forecasting Lab (NECEP)\\Portugal}

\date{\selectlanguage{english} \today}

\begin{document}

\maketitle

\selectlanguage{english}

\begin{abstract}

Partial least squares (PLS) is a simple factorisation method that works well with high dimensional problems in which the number of observations is limited given the number of independent variables. In this article, we show that PLS can perform better than ordinary least squares (OLS), least absolute shrinkage and selection operator (LASSO) and ridge regression in forecasting quarterly gross domestic product (GDP) growth, covering the period from 2000 to 2023. In fact, through dimension reduction, PLS proved to be effective in lowering the out-of-sample forecasting error, specially since 2020. For the period 2000-2019, the four methods produce similar results, suggesting that PLS is a valid regularisation technique like LASSO or ridge. \\\



\end{abstract}

\pagebreak


\pagebreak

\section{Introduction}
\label{sec:intro}

Economic shocks present challenges to forecasting and nowcasting exercises. All statistical models have problems, so the choice of a specific model or approach is always a balance among the risks of each model. The usage of ordinary least squares (OLS) when there is relevant information to the forecasting task is quite common. Two important motivations for the popularity of OLS are that i) it is a linear approximation to any true model; and ii) it uses all available statistical information for the task at hand. In the presence of economic shocks, however, the conditions for the robustness of OLS are not present, and one may be interested in other approaches that provide data-based forecasts that are more robust to different true relationships among predictor and predicted variables.

In fact, an OLS model could be overfitted to cope with dramatic changes in economic covariates. Thus, dealing with shocks requires simple models that provide small out-of-sample forecasting errors in a wide range of conditions. The procedure of reducing the complexity of statistical models to improve their predictability is called \emph{regularisation.}

In this paper we explore the usage of different regularisation techniques to forecast aggregated macroeconomic variables in the presence of a severe economic shocks, such as the COVID-19 pandemic and related lockdowns. Some popular regularisation techniques are ridge regression and least absolute shrinkage and selection operator (LASSO). They avoid overfitting by shrinking the least squares coefficients toward zero, or even set them to zero like LASSO, in a practical way to do variable selection \citep{Varian2014}.

Partial least squares (PLS) is a simple factorisation method introduced by \citet{Wold1966} that works well in statistical problems where predictors are large relative to the number of observations \citep{Taddy2019}. When the number of factors or directions in PLS converges to the number of predictor variables, PLS converges to OLS, provided the covariance matrix has full rank. Thus, a PLS model is a simplified OLS model subject to dimension reduction.

In practice, PLS is a regularisation technique to achieve dimension reduction, namely, in the context of big data \citep{Petropoulos2022}. The intuition is that some linear combination of the predictor variables, akin to building a scale, can achieve a good accuracy in a forecasting or nowcasting exercise. Additional orthogonal directions can be included if they help this exercise in a procedure called \emph{boosting}.

As it should be expected, OLS and regularised methods generate similar out-of-sample forecasting errors under normal conditions. However, methods that use regularisation techniques, such as ridge regression, LASSO or PLS, tend to perform better when severe economic shocks affect both the level and the statistical relationships among the variables \citep{Dauphin2022}.

A more surprising result is that PLS, even with a single direction, tends to perform better, in periods with severe shocks and potential changes of regime, than the other regularisation techniques studied. This is achieved without loss of performance during periods of more stable economic conditions. This surprising result highlights the practical usefulness and robustness of PLS, specially in demanding forecasting and nowcasting exercices.

Additionally, PLS can be applied on datasets with very few observations, even fewer than the number of variables. This feature allows for the rapid inclusion of new variables with short data histories, a common procedure used by macroeconomic forecasters during the COVID-19 pandemic.

The article is organised as follow: firstly, in section \ref{sec:literature}, we revise some relevant literature that inspired this research; then, in section \ref{sec:methods}, we describe succinctly the methods and techniques here adopted to forecast GDP, including cross-validation, regularisation and partial least squares; a brief description of the data and their treatment is made in section \ref{sec:data}; the next section presents the main findings for Portugal, Euro area and other countries, including the United States (section \ref{sec:findings}), followed by the conclusions and limitations of this work (section \ref{sec:conclusion}); additional figures and tables are presented in appendix.

\section{Related literature}
\label{sec:literature}

The COVID-19 lockdown recession highlighted the need for a systematic analysis of high-frequency indicators of economic activity \citep{Dauphin2022}. New indicators like Google search results have become available together with traditional high-frequency data series. For example, \citet{Woloszko2020} constructed the OECD Weekly Tracker of GDP growth from a quarterly model estimated on Google Trends (GT) search activities. A selection of 250 categories and 33 topics were adjusted for a common long-term trend with the \citet{Hodrick1997} filter. Seasonality and breaks were addressed too. Then, the relationship between GT variables and GDP growth was fitted using a feed-forward neural network that captures the non-linearities that are likely to be common in extreme situations like the COVID-19 pandemic.

The application of machine learning (ML) techniques like neural networks or PLS is a new feasible approach to nowcasting and forecasting. In a seminal paper, \citet{Varian2014} stresses that big datasets may require more powerful and flexible tools than linear regression and other conventional econometric techniques to capture complex relationships. In fact, the ability of summarising various sorts of nonlinear relationships is the main attractiveness of ML methods. Additionally, we may have more potential predictors than appropriate for estimation, so some kind of selection must be done. Here, penalised regressions like LASSO or PLS can save a lot of time from traditional methods to select the data variables with the strongest signals of economic activity. Similarly, \citet{Mullainathan2017} describe the supervised ML approach and its focus on the out-of-sample prediction performance, avoiding overfit to the training data through regularisation and parameter tuning.

Focused on real-time macroeconomic analysis, the article of \citet{Bok2018} reviewed how methods for tracking economic conditions using big data have evolved over time and explain how econometric techniques have advanced to mimic and automate the best practices of professional forecasters. They present in detail the methodology underlying the New York Fed Staff Nowcast, which employs these innovative techniques to produce early estimates of GDP growth, synthesising a wide range of macroeconomic data as they become available. These techniques include dynamic factor models (DFM) and vector autoregressions (VAR).

An exhaustive survey of methods and tools to forecast quarterly GDP from higher frequency indicators is provided by \citet{Dauphin2022}. Starting from literature on DFM estimated with the Kallman filter, they cover the recent literature that apply ML methods to economic data, namely, to nowcast GDP in countries like New Zealand, Indonesia, Lebanon or Turkey. Then, they apply ML methods (elastic net, ridge, LASSO, random forest, support vector machines, neural networks) side-by-side with a DFM and an autoregressive (AR) model of order 1 (benchmark) to a set of six European countries (Austria, Hungary, Ireland, Malta, Poland and Portugal).

For Portugal, \citet{Dauphin2022} found that a simple linear regression (OLS) lowers the mean absolute error (MAE) by 53.8\% from an AR model, but the ridge and LASSO regressions can lower that measure, respectively, by 15.2\% and 21.5\% from the OLS. In general, penalised regressions and support vector machines (SVM) perform well, but gains are less evident with random forests and neural networks. These results were obtained from a dataset for Portugal with 46 variables that covers the fields of national accounts, housing and construction, labour market, manufacturing, retail trade and consumption, international trade, financial indicatores, surveys and other (novel) data like Google searches for relevant keywords or air quality. The variable of interest was the quarterly year-on-year (y-o-y) GDP growth and the study covered a period of 23 quarters from the third quarter of 2015 to the first quarter of 2021, including the outliers related with the COVID-19 recessions and recoveries.  

According to the survey of \citet{Petropoulos2022}, the use of PLS in forecasting is common in finance applications where market returns are estimated from many predictors. Conventional OLS is highly susceptible to overfit in these applications, which is amplified by the typical noise in stock returns data. Thus, researchers have explored other methods, namely, based on principal component analysis. In this area, \citet{Huang2015} proposed a new investor sentiment index that aims to predict the aggregate stock market returns. Here, PLS was used to extract the most relevant common component from the sentiment proxies, separating out information that is relevant to the expected stock returns from the error and noise.

Following a similar approach, \citet{Kelly2013} used PLS to extract a single factor from a cross-section of portfolio-level book-to-market ratios in order to forecast the return and cash flow growth for the aggregate U.S. stock market. Their approach treats the challenging problem in empirical asset pricing: how does one exploit several predictors in relatively short time series? If the predictors number is near or more than the number of observations, the standard OLS regression may not work due to the fact that the data matrix may have a determinant of zero or close to it. This is the same kind of problem faced by economic forecasting during the COVID-19 pandemic, that can be tackled with PLS.

\section{Methods}
\label{sec:methods}

\subsection{Cross-validation}

In machine learning practice, it is common to divide the data into $k$ evenly sized random subsets, called \emph{folds}, for the purpose of training, testing and validation \citep{Varian2014} \citep[chapter 3]{Taddy2019}. Each one of these $k$-folds is used for testing the model fitted in the remaining $k-1$ subsets. The output is a set of $k$ predictions and associated loss than can be used to compute measures like the mean absolute error (MAE) or root mean squared error (RMSE), validating the predictive performance of the model on out-of-sample (OOS) data.     

\begin{figure}[!h]
	\centering
	\includegraphics[width=14cm]{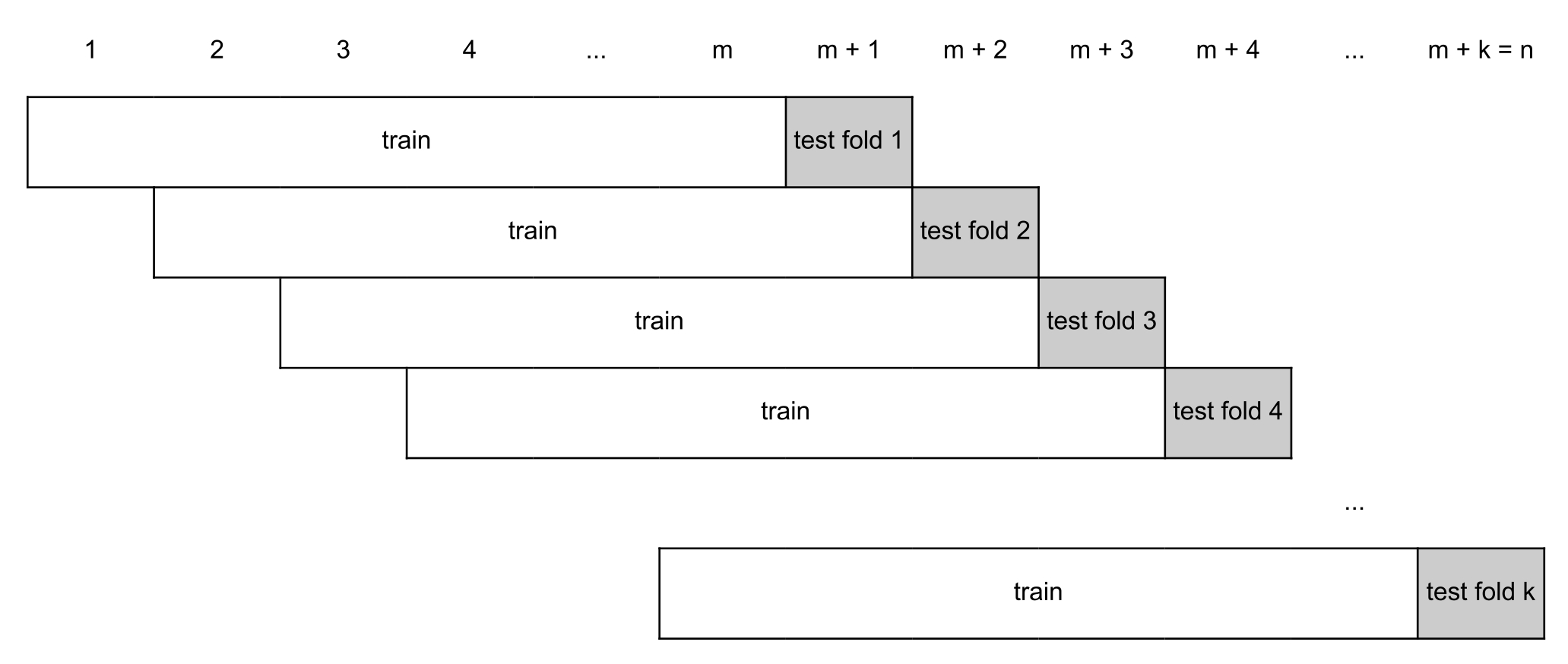}
	\caption{Cross validation adapted for time series}
	\label{fig:cv_ts}
\end{figure}

Cross-validation (CV) requires random sampling and independence between observations. Thus, its direct application to time series is limited in most cases because of the inherent serial correlation and potential non-stationarity of the data. In fact, standard $k$-fold CV is possible as long as the models considered have uncorrelated errors  \citep{Bergmeir2018}. In the general case, CV with time series must preserve training only with past observations, following an OOS evaluation scheme like the one represented in figure \ref{fig:cv_ts}, appendix. Here, the size of the training folds $m = n - k$, where $n$ is the total number of observations, is constant, following the classic CV for cross-sections. In the present application, a variable size of the training folds by fixing their origin in the first observation did not produce smaller forecast errors.

\subsection{Regularisation}

Complex models typically perform well in-sample rather than out-of-sample. Decision trees with a lot of terminal nodes (leafs) or regressions with several coefficients may capture not only the signal about the role of the inputs in predicting the response, but also fit the noise present in the training dataset, reducing the predictive power of these models with new samples \citep{Belloni2014}. Overfit can be avoided by reducing the dimension or complexity of the models, a procedure called \emph{regularisation}.

The regularisation of linear regressions can be performed by introducing a penalty as a function of a tuning parameter $\lambda$ in the least squares problem \citep{Belloni2014} \citep{Varian2014}:

\begin{equation}
	\hat{\beta} = \textrm{argmin} \left\{  \sum_{t=1}^{n} \left(y_t - \sum_{i=1}^{q} x_{it} \beta_i \right)^2 + \lambda \sum_{i=1}^{q} \left[ (1-\alpha)|\beta_i| + \alpha|\beta_i|^2\right]  \right\}.
	\label{eq:penreg}
\end{equation}

\noindent This problem with $0< \alpha <1$ is called \emph{elastic net regression}. If $\alpha = 1$, it is the \emph{ridge regression} that applies only a quadratic constraint on the OLS coefficients. Finally, if $\alpha = 0$, the problem becomes the above-mentioned LASSO. The last operator tend to shrink the coefficients towards zero, so it is particularly suitable for variable selection. As said, the LASSO can perform quite well with economic data because it provides a practical way to select the key variables among potential predictors, a sensible problem with high-dimensional data \citep{Belloni2014}, reducing the dimension of the model and favouring the OOS performance.

The penalty weight $\lambda$ in equation (\ref{eq:penreg}) is a signal-to-noise parameter that must be tuned to give a good signal with little noise \citep[chapter 3]{Taddy2019}. Cross-validation can be used once again to find the $\lambda$ that minimises the average OOS error (CV-min rule) or that gives an error no more than one standard error from that minimum (CV-1se rule). Alternatively, an information criteria like the Akaike's AIC can be used to measure that average error.

\subsection{Partial least squares}

\emph{Partial least squares} (PLS) is a supervised factorisation method that provides dimension reduction too with a wide range of applications in chemometrics, linguistics and finance. It is based on what \citet[chapter 7]{Taddy2019} calls the \emph{marginal regression} (MR), that is, a simple technique that constructs a single factor $z$ by summing up the predictors $x_{it}$ ($i=1, \dots, q$) for each observation $t$, weighted with the loadings $\varphi_{i}$ estimated with simple univariate regressions of $y$ on each $x_{i}$ independently. Then, the procedure fits a “forward" linear regression $y_t = \alpha + \beta z_t + \epsilon_t$ and computes the associated fitted values $\hat{y}_t = \hat{\alpha}+\hat{\beta} z_t$. The OLS slopes ($\varphi_{i}$ and $\beta$) are estimated simply by dividing the covariance between the response ($y$) and each independent variable (respectively, $x_i$ and $z$) by the variance of the latter.

The MR factor $z$ or $z^{1}$ is the first PLS direction. It is computed following the iterative algorithm \ref{alg:PLS} for $d=1$, starting from $\epsilon_t^{0} = y_t$. Then, the PLS algorithm takes the residuals $\epsilon_t^{1} = y_t - \hat{y}_t^{1}$ from the first MR, or PLS(1), and repeat a second MR to predict these residuals, updating the original estimate for $y$ by computing a second direction $z^{2}$ and estimating the corresponding “forward" regression. This procedure can be repeated once again by taking the residuals $\epsilon_t^{2} = y_t - \hat{y}_t^{2}$ from the second MR, or PLS(2), and so on, keeping in mind that the number of directions should be lower than or equal to the number of predictors $q$.

\begin{algorithm}
\caption{Partial Least Squares (PLS)}
\label{alg:PLS}
\begin{algorithmic}[1]
\State $\epsilon_t^{d-1} \gets y_t - \hat{y}_t^{d-1} \: \mathrm{with} \: \epsilon_t^{0} \gets y_t$
\State $\varphi_{i}^{d} \gets cov(x_i,\epsilon^{d-1})/var(x_i)$
\State $z_{t}^{d} \gets \sum_{i}^{}{x_{it} \varphi_{i}^{d}}$
\State $\beta_d \gets cov(z^{d},\epsilon^{d-1})/var(z^{d})$
\State $\hat{y}_t^{d} \gets \hat{y}_t^{d-1} + \beta_d z_t^{d} \: \mathrm{with} \: \hat{y}_t^{0} \gets \bar{y} - \beta_1 \bar{z}^{1}$
\State \Return $\varphi^{d}, z^{d}, \hat{y}^{d}$
\end{algorithmic}
\end{algorithm}

In practice, PLS reduces the high-dimensional matrix $x$ to a limited set of factors $z^{d}$ using the loadings or rotations $\varphi^{d}$ \citep{Taddy2013, Taddy2019}. Thus, PLS provides a parsimonious representation of $x$ by exploring the orthogonality between each factor $z^{d}$ and the residuals $\epsilon^{d} = y_t - \hat{y}_t^{d}$. By keeping the number of directions smaller than the number of predictors, PLS avoids the potential overfitting of a full model which is equivalent to OLS \citep{Sanchez2020}. In this sense, PLS is a regularisation technique like LASSO or ridge.

The construction of additive regression models by sequentially fitting a simple parameterised function to current residuals by least squares at each iteration, as PLS do, is called \emph{boosting} \citep{Friedman2001, Friedman2002}. This general procedure can result in dramatic improvements in OOS performance in many applications \citep{Friedman2000}. 

\pagebreak

\section{Data}
\label{sec:data}

In this application, we considered as response (dependent variable) the first difference of 100 times the natural log of Portuguese real GDP provided by Statistics Portugal (INE). Thus, every model was estimated in quarter-over-quarter (q-o-q) changes. In general, the seasonally adjusted predictors (independent variables) were submitted to the same transformation, with some exceptions indicated in table \ref{tab:predictors} (appendix). Similar datasets were built for the Euro area and its main economies (Germany, France, Italy and Spain), as well as for the United States. 

Data covered the period from the first quarter of 2000 to the fourth quarter of 2023 with a total of 95 q-o-q changes. The outliers concerned, namely, with the COVID-19 recessions and recoveries were treated with the Hampel filter. Using a sliding centred window of 19 quarters, this technique keeps the data around the median plus or minus 2.5 times the median absolute deviation (MAD), that is, the median of the absolute deviations from the median \citep{Leys2013}.

The Hampel filter is a robust technique because the median is very insensitive to the presence of outliers and the MAD is the natural (nonparametric) estimator of the “probable error" of a single observation \citep{Hampel1974}. In fact, the Hampel filter can reduce dramatically the difference between extreme points (maximum and minimum) in data distributions, as well as their standard deviations, but its impact is small on the mean in most cases. In practice, the Hampel filter changed only the 2020Q2 level of the Portuguese GDP, but the changes were more frequent among the predictors (independent variables). 

In the scope of the cross-validation procedure described in section \ref{sec:methods}, the number $k$ of testing folds was 36 (from 2015Q1 to 2023Q4) to assure that the performance measures (MAE and RMSE) would be computed with a pool of at least 30 out-of-sample predictions, that is, with a “large sample" from the statistical point of view. Thus, each training fold had 59 ($=95 - 36$) observations, starting from 2000Q2-2014Q4 to 2009Q1-2023Q3.

\section{Findings}
\label{sec:findings}

\subsection{Portugal}

Table \ref{tab:findings} presents the out-of-sample mean absolute (MAE) and root mean squared (RMSE) forecasting errors of the q-o-q change of the Portuguese GDP obtained with PLS models with 1, 2 and 3 directions side-by-side with some benchmarks, namely, an (integrated) autoregressive (AR) model of order 1, a linear (OLS) model, a ridge regression and a LASSO.

\begin{table}[h!]
	\caption{Mean OOS errors of the q-o-q log GDP difference by model (Portugal; training folds from 2000Q2-2014Q4 to 2009Q1-2023Q3; testing period: 2015Q1-2023Q4)} 
	\label{tab:findings} 
	\centering
	\begin{tabular}{lcc}
		  \hline
        Model                         & MAE   & RMSE    \\
	      \hline	
        Autoregressive (AR)           & 0.93  & 1.58  \\
		Linear regression (OLS)       & 0.60  & 1.00  \\
    	Ridge regression              & 0.58  & 0.97  \\
  		LASSO                         & 0.58  & 0.98  \\
        Marginal regression PLS(1)    & 0.57  & 0.96  \\
        PLS(2)                        & 0.58  & 0.97  \\
        PLS(3)                        & 0.59  & 0.97  \\
        Median of the best models     & 0.57  & 0.96  \\
        \hline
    \end{tabular}
\end{table}

Firstly, we found that the regression models perform better than the autoregressive (AR) benchmark. Thus, high frequency data may contain additional and relevant information to forecast the first difference of log GDP from the series itself.

Secondly, linear regression (OLS) performs reasonably well lowering the MAE by 35.3\% and the RMSE by 36.9\% from the AR model, but regularised regressions can perform even better. In particular, LASSO lowers the MAE by 4.2\% and the RMSE by 1.7\% from the OLS. In fact, LASSO selects only the relevant regressors, keeping industrial production, cement sales, retail trade turnover, card transactions,  services turnover, exports, imports, Eurozone Purchasing Managers' Index (PMI), employment and unemployment rate from a pool of 14 regressors at the end of the sample (table \ref{tab:coefficients2}, appendix). Importantly, LASSO changes the set of regressors as the sample window moves in time, allowing for an adaptive element in forecasting. The result was a regularised regression with few predictors than the original OLS regression that lowers the OOS forecast error. Ridge performs as LASSO in MAE and slightly better in RMSE.

Thirdly, partial least squares can perform even better than LASSO and ridge. In fact, we found that PLS with only one direction, that is, the marginal regression lowers the MAE by 6.1\% and the RMSE by 3.9\% from the OLS. Adding a second or third PLS direction did not improve that performance, probably due to overfitting. The regressors included in this analysis were selected on the expected relevance to predict GDP, so it is not surprising that they all capture the core economic activity. The surprising quality of PLS with a single direction suggests that most indicators are essentially capturing the underlying dynamics of the Portuguese economy.  

\begin{figure}[!h]
	\centering
	\includegraphics[width=14cm]{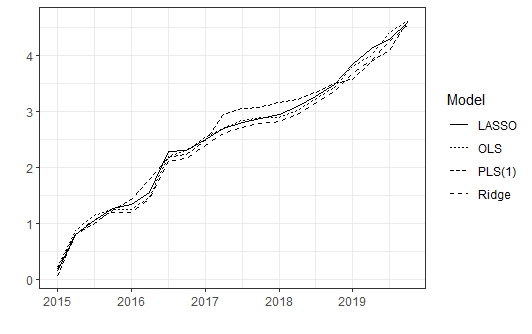}
	\caption{Cumulative absolute OOS error of the q-o-q log GDP difference by model  (Portugal; training folds from 2000Q2-2014Q4 to 2005Q1-2019Q3; testing period: 2015Q1-2019Q4)}
	\label{fig:error1}
\end{figure}

\begin{figure}[!h]
	\centering
	\includegraphics[width=14cm]{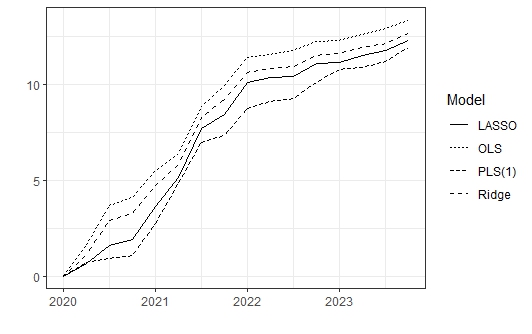}
	\caption{Cumulative absolute OOS error of the q-o-q log GDP difference by model  (Portugal; training folds from 2005Q2-2019Q4 to 2009Q1-2023Q3; testing period: 2020Q1-2023Q4)}
	\label{fig:error2}
\end{figure}

From tables \ref{tab:coefficients} and \ref{tab:coefficients2} (appendix), we can check that the PLS coefficients become closer to OLS ones by adding directions, but the regularised PLS(1) coefficients are more effective for forecasting purposes. In addition, figure \ref{fig:error1} reveals that the four methods (OLS, LASSO, ridge and PLS) produce similar cumulative absolute forecasting errors in “normal" times, that is, for the first testing period between 2015 and 2019, right before the COVID-19 pandemic. A different result was obtained for the second testing period (2020-2023), where PLS is undoubtedly the best model, as suggested by figure \ref{fig:error2}. Thus, the dimension reduction provided by PLS proved to be effective during the COVID-19 times and beyond, surpassing the performance of LASSO and ridge. The non-regularised model (OLS) got the greatest errors in the second period, possibly due to overfitting. Similar results were obtained for the original data, that is, without the treatment provided by the Hampel filter. 

\subsection{Euro area and other countries}

Similar results were obtained for the Euro area using the following predictors: industrial production (IPI), production in construction, retail trade turnover and Economic Sentiment Indicator (ESI). Here, PLS with only one direction (marginal regression) performs better than LASSO, ridge and PLS with two or three directions (see table \ref{tab:euro_area}). Also, we found that ridge may perform better than LASSO.

\begin{table}[h!]
	\caption{Mean absolute OOS error of the q-o-q log GDP difference by model (Euro area, Germany, France, Italy,  Spain and United States; training folds: from 2000Q2-2014Q4 to 2009Q1-2023Q3; testing period: 2015Q1-2023Q4)} 
	\label{tab:euro_area} 
	\centering
	\begin{tabular}{lcccccc}
		  \hline
        Model                         & Euro area & Germany & France & Italy & Spain & U.S.\\
	      \hline	
        Autoregressive (AR)           & 0.48 & 0.51 & 0.62 & 0.60 & 0.60 & 0.76\\
		Linear regression (OLS)       & 0.45 & 0.46 & 0.57 & 0.48 & 0.62 & 0.52\\
    	Ridge regression              & 0.43 & 0.45 & 0.55 & 0.46 & 0.58 & 0.49\\
  		LASSO                         & 0.45 & 0.45 & 0.56 & 0.48 & 0.59 & 0.48\\
        Marginal regression PLS(1)    & 0.40 & 0.43 & 0.48 & 0.43 & 0.49 & 0.52\\
        PLS(2)                        & 0.45 & 0.46 & 0.55 & 0.44 & 0.54 & 0.47\\
        PLS(3)                        & 0.45 & 0.46 & 0.57 & 0.46 & 0.59 & 0.47\\
        Median of the best models     & 0.44 & 0.46 & 0.56 & 0.47 & 0.58 & 0.49\\
        \hline
    \end{tabular}
\end{table}
 
The same method was applied to the four biggest economies of the Euro area (Germany, France, Italy and Spain) adding PMI, exports and imports to the previous pool of predictors (see the same table). PLS(1) is always the best technique, given small mean absolute errors specially for Germany and Italy. The ridge regression and the median of the best models performs quite well too, following the results obtained for the Euro area and Portugal.

\begin{figure}[!p]
	\centering
	\includegraphics[width=14cm]{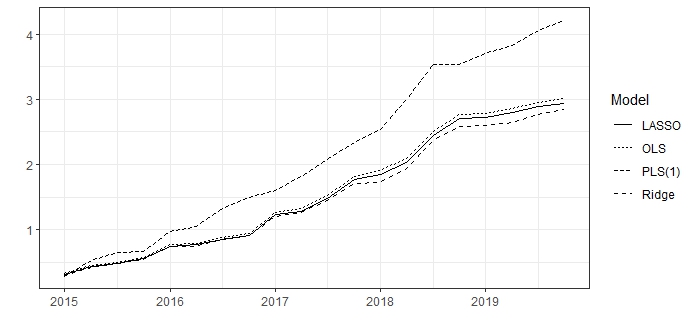}
	\caption{Cumulative absolute OOS error of the q-o-q log GDP difference by model  (Euro area; training folds from 2000Q2-2014Q4 to 2005Q1-2019Q3; testing period: 2015Q1-2019Q4)}
	\label{fig:error_EA1}
\end{figure}

\begin{figure}[!h]
	\centering
	\includegraphics[width=14cm]{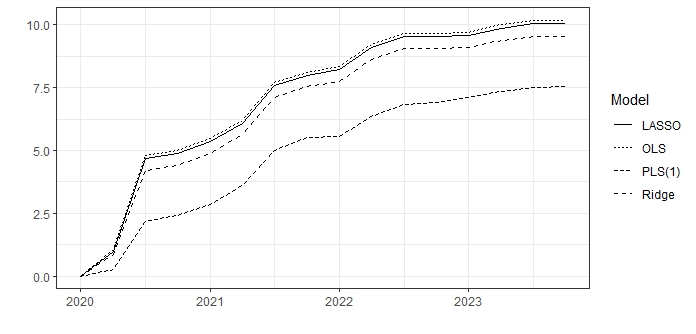}
	\caption{Cumulative absolute OOS error of the q-o-q log GDP difference by model  (Euro area; training folds from 2005Q2-2019Q4 to 2009Q1-2023Q3; testing period: 2020Q1-2023Q4)}
	\label{fig:error_EA2}
\end{figure}

\begin{figure}[!h]
	\centering
	\includegraphics[width=14cm]{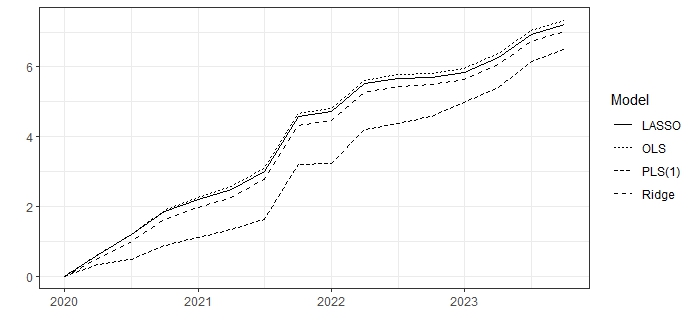}
	\caption{Cumulative absolute OOS error of the q-o-q log GDP difference by model  (Germany; training folds from 2005Q2-2019Q4 to 2009Q1-2023Q3; testing period: 2020Q1-2023Q4)}
	\label{fig:error_GE2}
\end{figure}

\begin{figure}[!h]
	\centering
	\includegraphics[width=14cm]{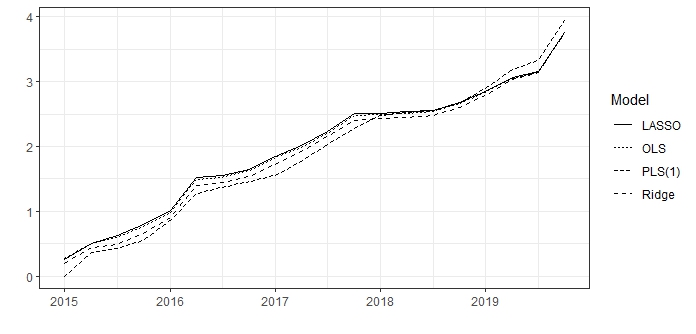}
	\caption{Cumulative absolute OOS error of the q-o-q log GDP difference by model  (France; training folds from 2000Q2-2014Q4 to 2005Q1-2019Q3; testing period: 2015Q1-2019Q4)}
	\label{fig:error_FR1}
\end{figure}

\begin{figure}[!h]
	\centering
	\includegraphics[width=14cm]{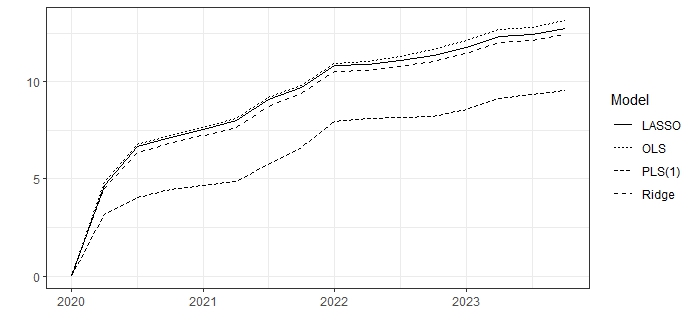}
	\caption{Cumulative absolute OOS error of the q-o-q log GDP difference by model  (France; training folds from 2005Q2-2019Q4 to 2009Q1-2023Q3; testing period: 2020Q1-2023Q4)}
	\label{fig:error_FR2}
\end{figure}

\begin{figure}[!h]
	\centering
	\includegraphics[width=14cm]{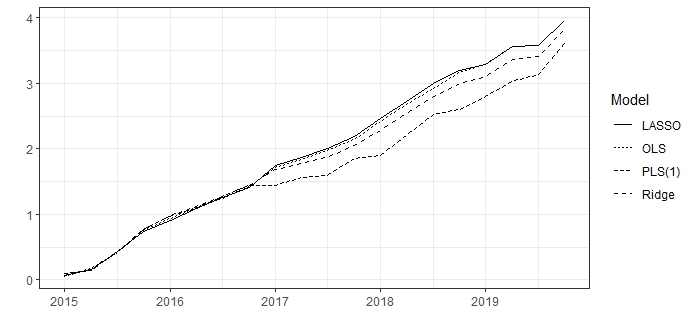}
	\caption{Cumulative absolute OOS error of the q-o-q log GDP difference by model  (Italy; training folds from 2000Q2-2014Q4 to 2005Q1-2019Q3; testing period: 2015Q1-2019Q4)}
	\label{fig:error_IT1}
\end{figure}

\begin{figure}[!h]
	\centering
	\includegraphics[width=14cm]{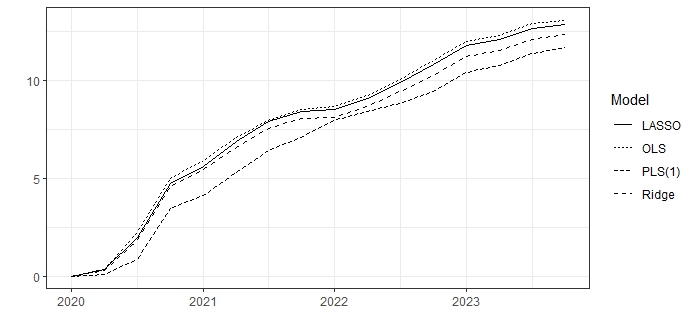}
	\caption{Cumulative absolute OOS error of the q-o-q log GDP difference by model  (Italy; training folds from 2005Q2-2019Q4 to 2009Q1-2023Q3; testing period: 2020Q1-2023Q4)}
	\label{fig:error_IT2}
\end{figure}

\begin{figure}[!h]
	\centering
	\includegraphics[width=14cm]{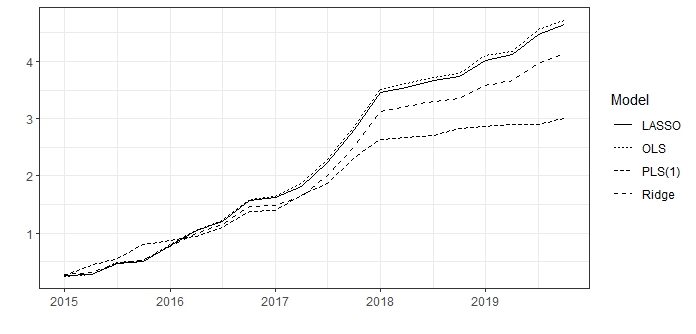}
	\caption{Cumulative absolute OOS error of the q-o-q log GDP difference by model  (Spain; training folds from 2000Q2-2014Q4 to 2005Q1-2019Q3; testing period: 2015Q1-2019Q4)}
	\label{fig:error_SP1}
\end{figure}

\begin{figure}[!h]
	\centering
	\includegraphics[width=14cm]{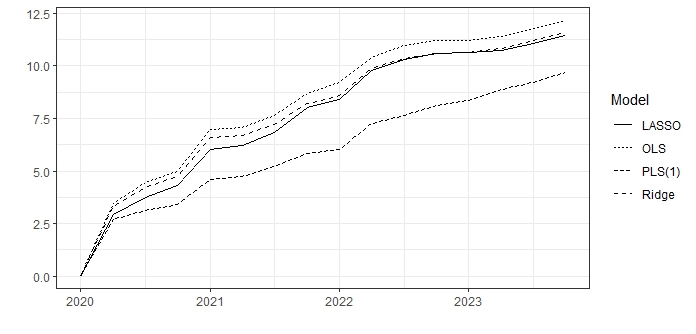}
	\caption{Cumulative absolute OOS error of the q-o-q log GDP difference by model  (Spain; training folds from 2005Q2-2019Q4 to 2009Q1-2023Q3; testing period: 2020Q1-2023Q4)}
	\label{fig:error_SP2}
\end{figure}

Figures \ref{fig:error_EA2}, \ref{fig:error_GE2}, \ref{fig:error_FR2}, \ref{fig:error_IT2} and \ref{fig:error_SP2} (appendix) suggest that PLS performs better in the European countries than the other methods (OLS, LASSO and ridge) in terms of cumulative absolute OOS error since 2020, in coherence with the findings for Portugal above described. Even before 2020, PLS can perform better as suggested by the cases of Italy (figure \ref{fig:error_IT1}) and Spain (figure \ref{fig:error_SP1}). Nevertheless, for the Euro area (figure \ref{fig:error_EA1}) and France (figure \ref{fig:error_FR1}), PLS was the worst method between 2015 and 2019, accumulating the greater OOS error. 

\begin{figure}[!h]
	\centering
	\includegraphics[width=14cm]{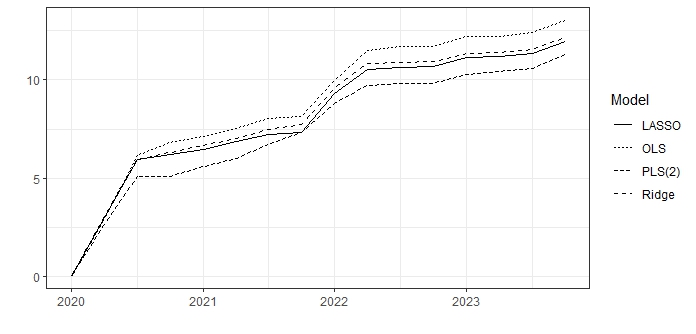}
	\caption{Cumulative absolute OOS error of the q-o-q log GDP difference by model  (United States; training folds from 2005Q2-2019Q4 to 2009Q1-2023Q3; testing period: 2020Q1-2023Q4)}
	\label{fig:error_US2}
\end{figure}

For the United States, we found that PLS with two directions performs better than LASSO, ridge and OLS, specially since 2020 (figure \ref{fig:error_US2}). This result was obtained using industrial production, retail trade sales, University of Michingan's consumer sentiment, all employment (non-farm), export and imports as predictors.

\section{Conclusions and limitations}
\label{sec:conclusion}

The main conclusion of this article is that PLS can improve dramatically the out-of-sample performance of forecasting models traditionally estimated with autoregressions or OLS. This is a rather simple way to improve the quality of GDP forecasts that may be favoured over regularisation techniques like LASSO or ridge regressions, following the results obtained for the United States, Euro area, Germany, France, Italy, Spain and Portugal, specially for the period 2020-2023.

The remarkable performance of PLS seems to be related with its own way to reduce the dimension of the problem by introducing a single factor, avoiding overfitting, namely, in a period characterised by important shifts in GDP growth with the COVID-19 recessions and recoveries.

Naturally, the results obtained could be somewhat different using vintage time series that unwittingly can introduce endogeneity to the out-of-sample forecasting exercise. In addition, we used complete information concerned with the three months of each quarter, instead of incomplete information for the current quarter like in real-time applications. In fact, adding information on the current quarter can reduce dramatically the forecasting errors, as shown in a previous article \citep{Assuncao2022a}. Thus, a real-time nowcasting exercise might give greater errors and, eventually, a different rank among the regularised models.

Another limitation of this work is that it explored only linear regularisation techniques like ridge, LASSO and PLS. In general, tools like neural networks may improve the out-of-sample forecasting performance, specially with big data and in the presence of economic shocks and changes of regime, due to the possible non-linearity of these phenomena.

Further research may include the development of a multivariate PLS model where the dependent variables will be the GDP and its components, following the approach of \citet{Abdi2003}.

\section*{Acknowledgments}
\label{sec:acknowledgments}

This work was financed by Fundação para a Ciência e a Tecnologia (FCT) under a doctorate auxiliary researcher grant at Universidade Católica Portuguesa (UCP) - Católica Lisbon Research Unit in Business \& Economics (CUBE) with the digital object identifier (DOI): \href{https://doi.org/10.54499/CEECINST/00070/2021/CP1778/CT0008}{10.54499/CEECINST/00070/2021/CP1778/CT0008}.

\pagebreak

\section*{Abbreviations}
\label{sec:abbreviations}

The following abbreviations are used in this paper:\\

\noindent 
\begin{tabular}{@{}ll}
ACAP    & Associação Automóvel de Portugal\\ 
AIC     & Akaike's Information Criterion\\
AR      & Autoregressive\\
ATM     & Automated Teller Machines\\
BdP     & Banco de Portugal\\
CV      & Cross-Validation\\
DFM     & Dynamic Factor Models\\
EC      & European Commission\\
ESI     & Economic Sentiment Indicator\\
FCT     & Fundação para a Ciência e a Tecnologia\\
GDP     & Gross Domestic Product\\
GT      & Google Trends\\ 
INE     & Statistics Portugal\\
IPI     & Industrial Production Index\\
LASSO   & Least Absolute Shrinkage and Selection Operator\\
MAE     & Mean Absolute Error\\
MAD     & Median Absolute Deviation\\
MF      & Portuguese Government - Ministry of Finance\\
ML      & Machine Learning\\
MR      & Marginal Regression\\
MSE     & Mean Squared Error\\
OECD    & Organisation for Economic Co-operation and Development\\
OLS     & Ordinary Least Squares\\
OOS     & Out-Of-Sample\\
PLS     & Partial Least Squares\\
PMI     & Purchasing Managers’ Index\\
POS     & Point of Sales\\
q-o-q   & quarter-over-quarter\\
RMSE    & Root Mean Squared Error\\
S\&P    & Standard and Poors\\
SVM     & Support Vector Machines\\
U.S.    & United States\\
VAR     & Vector Autoregressions\\
y-o-y   & year-over-year
\end{tabular}

\pagebreak

\bibliography{library}

\pagebreak

\section*{Appendix}
\label{sec:appendix}

\begin{table}[p]
	\caption{List of predictors and their sources and transformations (Portugal)} 
	\label{tab:predictors} 
	\centering
	\begin{tabular}{lccc}
	    \hline	
        Variable                              & Source  & Transformation  & q-o-q difference\\
	    \hline	
        Industrial Production Index (IPI)     & INE     & 100 $\times$ log & Yes\\
		Cement sales    				      & MF	    & 100 $\times$ log & Yes\\
    	Retail trade turnover			      & INE	    & 100 $\times$ log & Yes\\
        Card transactions at POS/ATM	      & BdP	    & 100 $\times$ log & Yes\\
		Car sales          					  & ACAP    & 100 $\times$ log & Yes\\
    	Van sales          					  & ACAP    & 100 $\times$ log & Yes\\
        Bus and truck sales					  & ACAP    & 100 $\times$ log & Yes\\
        Services turnover                     & INE     & 100 $\times$ log & Yes\\
        Exports of goods                      & INE     & 100 $\times$ log & Yes\\
        Imports of goods                      & INE     & 100 $\times$ log & Yes\\
        Economic Sentiment Indicator (ESI)    & EC	    & Level - 100      & No\\
        Eurozone Purchasing Managers' Index (PMI) & S\&P& Level - 50       & No\\
        Employment                            & INE     & 100 $\times$ log & Yes\\
        Unemployment rate                     & INE     & \%               & Yes\\
        \hline
    \end{tabular}
\end{table}

\begin{table}[h]
	\caption{Descriptive statistics (Portugal; q-o-q differences except the variables indicated with *)} 
	\label{tab:desc_qoq} 
	\centering
	\begin{tabular}{lcccc}
		  \hline
        Variables               & Mean     & Standard Dev.   & Min     & Max   \\
	      \hline	        GDP                     & 0.2      & 1.2     & -4.5    & 4.4   \\
        IPI                     & -0.1     & 1.9     & -5.2    & 7.3   \\
		Cement sales    		& -1.1     & 6.2     & -21.6   & 10.0  \\
		Retail trade turnover	& 0.1      & 1.7     & -6.1    & 7.1   \\
        Card transactions	    & 1.1      & 4.5     & -17.9   & 17.8  \\
        Car sales          		& -0.6     & 14.6    & -55.6   & 60.9  \\
        Van sales          		& -1.5     & 16.0    & -72.5   & 36.3  \\
        Bus and truck sales		& 0.0      & 20.6    & -49.1   & 71.6  \\
        Services turnover       & -0.1     & 2.3     & -6.3    & 5.4   \\
        Exports of goods        & 0.9      & 3.2     & -8.7    & 7.3   \\
        Imports of goods        & 0.7      & 2.9     & -8.9    & 5.7   \\
        ESI* (Level - 100)      & 0.4      & 8.9     & -20.7   & 13.9  \\
        PMI* (Level - 50)       & 2.6      & 3.6     & -3.9    & 10.5  \\
        Employment              & 0.1      & 0.7     & -2.5    & 1.9   \\
        Unemployment            & 0.0      & 0.4     & -1.0    & 1.1   \\
        \hline   
    \end{tabular}
\end{table}

\begin{table}[!t]
	\caption{Estimated coefficients for the first 59 quarters of the sample (Portugal, 2000Q2-2014Q4)} 
	\label{tab:coefficients} 
	\centering
	\begin{tabular}{lcccc}
		  \hline
        Predictors                           & OLS       & LASSO  & PLS(1)   & PLS(6)\\
	      \hline	
        Industrial Production Index (IPI)    & 0.066     & 0.055  & 0.042    & 0.067 \\
		Cement sales    				     & 0.048***  & 0.043  & 0.017    & 0.047 \\
		Retail trade turnover			     & 0.068     & 0.056  & 0.070    & 0.066 \\
        Card transactions at POS/ATM	     & 0.009     & -      & 0.011    & 0.009 \\
        Car sales          					 & 0.015     & 0.010  & 0.006    & 0.015 \\
        Van sales          					 & 0.005     & 0.005  & 0.004    & 0.005 \\
        Bus and truck sales					 & 0.003     & 0.002  & 0.002    & 0.004 \\
        Services turnover                    & 0.010     & 0.012  & 0.046    & 0.011 \\
        Exports of goods                     & 0.012     & -      & 0.007    & 0.011 \\
        Imports of goods                     & 0.002     & 0.017  & 0.028    & 0.004 \\
        Economic Sentiment Indicator (ESI)   & 0.016     & 0.016  & 0.012    & 0.015 \\
        Eurozone Purchasing Managers' Index (PMI)& 0.028 & 0.019  & 0.021    & 0.027 \\
        Employment                           & -0.027    & 0.012  & 0.144    & 0.012 \\
        Unemployment rate                    & -0.108    & -0.075 & -0.208   & -0.050 \\
        Constant                             & 0.208*    & 0.222  & 0.164    & 0.206 \\
        \hline
        \multicolumn{5}{l}{\small{\emph{p}-values: * $p<0.10$; ** $p<0.05$; *** $p<0.01$.}}
    \end{tabular}
\end{table}

\pagebreak

\begin{table}[h]
	\caption{Estimated coefficients for the last 59 quarters of the sample (Portugal, 2009Q2-2023Q4)} 
	\label{tab:coefficients2} 
	\centering
	\begin{tabular}{lcccc}
        \hline
		  Predictors                           & OLS       & LASSO  & PLS(1)   & PLS(6)\\
	      \hline	
        Industrial Production Index (IPI)    & 0.117*    & 0.087  & 0.071    & 0.121 \\
		Cement sales    				     & 0.034*    & 0.012  & 0.008    & 0.034 \\
		Retail trade turnover			     & 0.171*    & 0.162  & 0.106    & 0.171 \\
        Card transactions at POS/ATM	     & 0.059     & 0.088  & 0.049    & 0.075 \\
        Car sales          					 & 0.001     & -      & 0.008    & 0.000 \\
        Van sales          					 & 0.002     & -      & 0.004    & 0.004 \\
        Bus and truck sales					 & 0.000     & -      & 0.001    & -0.001 \\
        Services turnover                    & 0.183***  & 0.124  & 0.082    & 0.186 \\
        Exports of goods                     & 0.070     & 0.058  & 0.036    & 0.072 \\
        Imports of goods                     & -0.099**  & -0.041 & 0.035    & -0.108 \\
        Economic Sentiment Indicator (ESI)   & -0.021    & -      & 0.011    & -0.020 \\
        Eurozone Purchasing Managers' Index (PMI)& 0.078*& 0.048  & 0.041    & 0.071 \\
        Employment                           & 0.547**   & 0.314  & 0.206    & 0.471 \\
        Unemployment rate                    & 0.719*    & 0.403  & -0.109   & 0.597 \\
        Constant                             & 0.104     & 0.083  & 0.053    & 0.118 \\
        \hline
        \multicolumn{5}{l}{\small{\emph{p}-values: * $p<0.10$; ** $p<0.05$; *** $p<0.01$.}}
    \end{tabular}
\end{table}

\begin{figure}[!h]
	\centering
	\includegraphics[width=14cm]{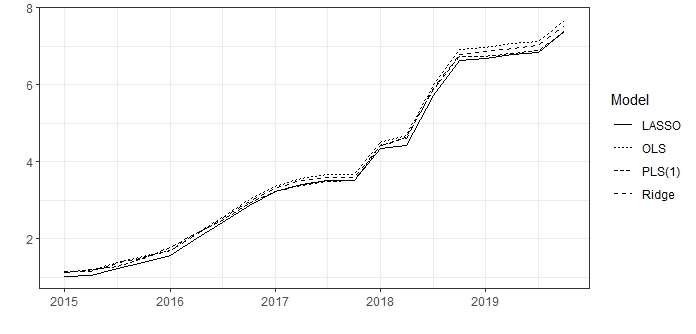}
	\caption{Cumulative absolute OOS error of the q-o-q log GDP difference by model  (Germany; training folds from 2000Q2-2014Q4 to 2005Q1-2019Q3; testing period: 2015Q1-2019Q4)}
	\label{fig:error_GE1}
\end{figure}

\begin{figure}[!h]
	\centering
	\includegraphics[width=14cm]{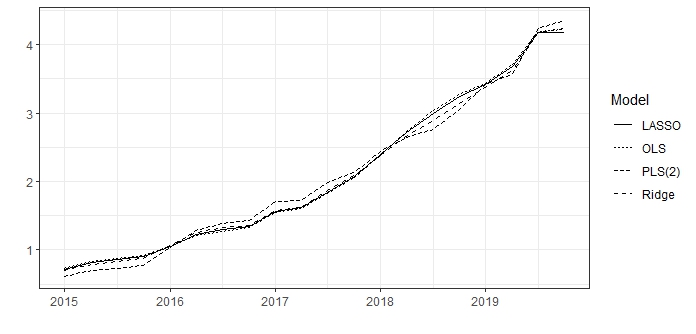}
	\caption{Cumulative absolute OOS error of the q-o-q log GDP difference by model  (United States; training folds from 2000Q2-2014Q4 to 2005Q1-2019Q3; testing period: 2015Q1-2019Q4)}
	\label{fig:error_US1}
\end{figure}

\end{document}